\documentclass[a4paper,11pt]{article}
\usepackage{pos}
\def\be{\begin{equation}}
\def\ee{\end{equation}}
\def\Rbar{\overline R}
\def\imp{\Rightarrow}
\def\nn{\nonumber\\}
\def\pa{\partial}
\def\Scal{{\cal {S}}}
\def\msbar{{\overline{\rm MS}}}
\def\momp{{$\hbox{MOM}^{\prime}$ }}
\usepackage{tikz}
\usetikzlibrary{calc}
\usetikzlibrary{shapes}
\usetikzlibrary{arrows,shapes,positioning}
\usetikzlibrary{decorations.markings}
\tikzstyle arrowstyle=[scale=1]
\tikzstyle directed=[postaction={decorate,decoration={markings,
    mark=at position .5 with {\arrow[arrowstyle]{stealth}}}}]
\tikzstyle reverse directed=[postaction={decorate,decoration={markings,
    mark=at position .5 with {\arrowreversed[arrowstyle]{stealth};}}}]

\usetikzlibrary{intersections}
\usetikzlibrary{calc}
\usetikzlibrary{shapes}

\newcommand{\vcross}{\node[draw=black, fill,line width=0.25mm,cross out,inner sep=5pt,minimum size=0pt]}

\renewcommand{\logo}{\relax}

\title{No-$\pi$ schemes for multi-coupling theories}

\author*[a]{Ian Jack}
\renewcommand{\printHeadAuthors}{Ian Jack}
\affiliation[a]{Theoretical Physics Division, Department of Mathematical Sciences, University of Liverpool, P.O. Box 147,
  Liverpool L69 3BX, UK}

\emailAdd{dij@liverpool.ac.uk}

\abstract{We show that even $\zeta$-functions may be removed from the $\beta$-functions of general multi-coupling theories up to high loop order by means of coupling redefinitions. For theories whose $\beta$-function is determined by the anomalous dimensions of the fields, such as supersymmetric theories, this corresponds to a renormalisation scheme change to a momentum subtraction scheme.}

\FullConference{Loops and Legs in Quantum Field Theory (LL2024)\\
 14-19 April, 2024\\
Wittenberg, Germany\\}

\begin{document}
\begin{flushright}
{\bf LTH1375}
\end{flushright}
\date{}
\vspace*{3mm}
\maketitle
\section{Introduction}

It was noticed some time ago that $\zeta_4$ cancelled in the QCD Adler function up to $ O(\alpha_s^3)$\cite{GKL} and $ O(\alpha_s^4)$ \cite{baikov6,BCK1}.
 Further empirical support for even-$\zeta$ cancellation was provided for instance in Refs.~\cite{davies,herz1,herz2}.
The phenomenon was explained at 3 loops \cite{david} and 4 loops \cite{baikov4,BCK2},  in terms of the dependence of Feynman integrals on certain combinations of $ \zeta$-functions, such that even $\zeta$s (which may be expressed as powers of $\pi^2$) only appeared in a well-defined way in conjunction with odd $\zeta$s.  However, it was remarked in Refs.~\cite{baikov5}, \cite{herz} that the $\zeta_4$-dependence in the scalar quark and scalar gluonium correlators respectively did not cancel at $O(\alpha_s^4)$, at least in standard minimal subtraction. On the other hand, in a parallel development, a so-called ``$ C$-scheme" was introduced\cite{boito}, in which it was observed that additional RG functions were free of even $ \zeta$s up to certain orders\cite{jamin}. Soon afterwards it was proposed that the "no-$\pi$" property holds to all orders in the scheme termed the $\hat G$ scheme\cite{baikov1,baikov2,baikov3}.
 
Meanwhile work on momentum subtraction (MOM) schemes has also demonstrated evidence for the no-$\pi$ property\cite{chetret1,vonsmek,gracey1,gracey2,gracey3}. All the developments so far mentioned have been for single-coupling theories except in the case of the supersymmetric Wess-Zumino model, for which the no-$\pi$ property was proved up to five loops for a general tensor coupling\cite{gracey4}. 

In this talk we present a proposal (described in more detail in Ref.~\cite{jack}) which extends the earlier ideas described here to the multicoupling case and also has the potential to set the various schemes mentioned above in a unified context. We suggest that for any (multi-coupling) theory there are at least two, and possibly several, renormalisation schemes in which even $ \zeta$s are absent. These can all be specified by redefinition of the couplings. Some of these schemes have simple physical definitions. One of them is a minimal scheme where we absorb even-$ \zeta$-dependent finite parts of divergent $n$-point functions; another is a variant of "MOM", i.e. it is specified by absorbing all finite parts of $n$-point functions in some well-defined way. It is easiest to demonstrate these suggestions for the supersymmetric Wess-Zumino model where the $\beta$ function is defined by the two-point function, but we believe that the idea has a wider application. The immediate consequence of this choice of model is that the interactions amongst a multiplet of $N$ superfields $\Phi_i$, $i=1,\ldots N$, are defined by the superpotential 
\be
W(\Phi)=g^{ijk}\Phi_i\Phi_j\Phi_k+\hbox{c.c.},
\label{Wdef}
\ee leading to three-point interactions; so we shall make this simplifying assumption in the rest of the talk.

\section{Basic ideas}

We start by describing the basic features of  the subtraction procedure in order to establish notation. We define counterterms recursively by the standard $ R$-operation. We define the counterterm $F$ by 
\be
F=\sum F(G), \quad F(G)=-\Rbar(G).
\label{rbar}
\ee
Here the sum is over all the relevant graphs $ G$ (in this case the two-point graphs with three-point vertices which contribute to the anomalous dimension), and $\Rbar$ denotes subtractions of diagrams with counterterm insertions corresponding to divergent subgraphs. To give a simple two-loop example, for the single two-loop graph $G_2$ contributing to the anomalous dimension we have
\be
\Rbar(G_2)= 
\tikz[baseline=(vert_cent.base)]{
  \node (vert_cent) {\hspace{-13pt}$\phantom{-}$};
  \draw (-0.4,0)--(0.1,0)
       (0.7,0) ++(0:0.6cm and 0.4cm) arc (0:50:0.6cm and 0.4cm) node (n1)
        {}
       (0.7,0) ++(50:0.6cm and 0.4cm) arc (50:130:0.6cm and 0.4cm) node (n2) 
       {}
       (0.7,0) ++(130:0.6cm and 0.4cm) arc (130:360:0.6cm and 0.4cm)
        (n1.base) to[out=215,in=325] (n2.base)
                (1.3,0)--(1.8,0);
          }
-\frac{F_{1,1}}{\epsilon}
\tikz[baseline=(vert_cent.base)]{
  \node (vert_cent) {\hspace{-13pt}$\phantom{-}$};
  \draw (-0.4,0)--(0.1,0)
        (0.7,0) ++(0:0.6cm) arc (0:360:0.6cm and 0.4cm)
        (1.3,0)--(1.8,0);
\vcross at (0.1,0){};},
\label{diag}
\ee
where $F_{1,1}$ is the one-loop simple pole contribution to $F$ (shortly to be defined in general). We are using here a somewhat schematic notation in which the diagram stands both for the Feynman integral and the associated product of couplings. 
We write $d=4-\epsilon$ so that divergences appear as poles in $\epsilon$, and denote the usual dimensional regularisation mass parameter by $\mu$. We  include finite parts in the definition of $F$ in Eq.~\eqref{rbar}; though standard minimal subtraction involves subtraction only of the pole term.
We write the bare coupling $ g_B$ (assuming now the use of  minimal subtraction) as
\be
g_B=\mu^{\frac12\epsilon}\left(g+\frac{1}{\epsilon}\left\{F_{1,1}+F_{2,1}+F^{\zeta_3}_{3,1}\zeta_3+F^{\zeta_4}_{4,1}\zeta_4\right\}+\frac{1}{\epsilon^2}\left\{F_{2,2}+F^{\zeta_3}_{4,2}\zeta_3\right\}\right)+\ldots
\ee
Here, $ F^{\zeta_n}_{L,m}$ denotes the $ L$-loop, order $ \epsilon^{-m}$ $ \zeta_n$-dependent contribution to $ F$, while $ F_{L,m}$ similarly denotes the $ L$-loop, order $ \epsilon^{-m}$ non-$ \zeta$-dependent contribution to $ F$.
We have
\be
\mu\frac{d}{d\mu} g_B=0,
\label{gbare}
\ee
and the $ \beta$-function is defined by
\be
\hat\beta(g)=\mu\frac{d}{d\mu}g=-\epsilon g+\beta(g)
\ee
which entails
\be
\beta_L^{\zeta_n}=LF^{\zeta_n}_{L,1}, \quad \beta_L=LF_{L,1},
\label{betas}
\ee
where $ \beta_L^{\zeta_n}$, $ \beta_L$ are the $ \zeta_n$-dependent and purely rational  $ L$-loop contributions, respectively, to $ \beta(g)$. We also have from Eq.~\eqref{gbare}
\be
F^{\zeta_3}_{4,2}=\frac14\left(\beta^{\zeta_3}_{3}\cdot F_{1,1}+\beta_1\cdot F^{\zeta_3}_{3,1}\right)=\frac14\left(3F^{\zeta_3}_{3,1}\cdot F_{1,1}+F_{1,1}\cdot F^{\zeta_3}_{3,1}\right),
\label{double}
\ee
where
\be 
f\cdot\equiv f\frac{\pa}{\pa g}\quad \hbox{or}\quad f\cdot\equiv f^{ijk}\frac{\pa}{\pa g^{ijk}}
\ee
depending on whether we have a single coupling $g$ or a tensor coupling $g^{ijk}$. In minimal subtraction the first appearance of $\zeta_4$ is at four loops, so 
 we now want to find a renormalisation scheme where $ \beta^{\zeta_4}_4=0$.
 A change of scheme corresponds to a coupling redefinition $ g\rightarrow g'(g)$ which leads to a variation $ \beta(g)\rightarrow \beta'(g')$ given by 
\be
\beta'(g')=\mu\frac{d}{d\mu}g'=(\beta(g))^{klm}\frac{\pa}{\pa g^{klm}} g'(g)=\beta(g)\cdot  g'(g).
\ee
An infinitesimal variation $ g'=g+\delta g$ then gives $ \beta'=\beta+\delta\beta$ with
\be
\delta\beta=[\beta,\delta g]+\ldots
\label{delb}
\ee
with the commutator defined by
\be
 [X,Y]=X\cdot Y-Y\cdot X.
\ee
\section{Properties of Feynman diagrams}
In this Section we give a brief introduction to the properties of Feynman diagrams with regard to even $\zeta$-functions which lead to the existence of schemes in which these even $\zeta$s are absent. It turns out that the even-$ \zeta$ terms in (at least a large class of) Feynman diagrams are not independent; in fact they only occur in certain combinations with the odd $\zeta$-functions.  For an illustration, consider the simple generic loop
\be
\tikz[baseline=(vert_cent.base)]{
  \node (vert_cent) {\hspace{-13pt}$\phantom{-}$};
\draw[directed] (-0.4,0)--(0.1,0);
    \draw (0.7,0) ++(0:0.6cm and 0.4cm) arc (0:180:0.6cm and 0.4cm) node(n1) {}
             (0.7,0) ++(180:0.6cm and 0.4cm) arc (180:360:0.6cm and 0.4cm) node(n2){};
             \draw (1.3,0)--(1.7,0); 
\draw (-.2,0) node[above]{\tiny{$p_{\mu}$}}; 
\draw (.6,.3) node[above]{\tiny{$a$}}; 
\draw (.6,-0.4) node[above]{\tiny{$b$}}; 
}
\ee
(where $a$, $b$ are the powers of the propagators and $p_{\mu}$ is the incoming momentum) for which we have the well-known result
\be
\frac{1}{(p^2)^{a+b-\frac{d}{2}}}L(a,b),\quad \hbox{where}\quad L(a,b)=(4\pi)^{\frac{\epsilon}{2}}\frac{\Gamma(\frac{d}{2}-a)\Gamma(\frac{d}{2}-b)\Gamma(a+b-\frac{d}{2})}{\Gamma(a)\Gamma(b)\Gamma(d-a-b)}.
\ee
A large class of Feynman integrals may be written as a product of $ L(a,b)$ for various $ a$, $ b$. For logarithmically divergent integrals the arguments $a$, $b$ are of the form $a=1+\alpha\epsilon$, $b=1+\beta\epsilon$.
There is an expansion
\begin{align}
L(1+\alpha\epsilon,1+\beta\epsilon)=&\frac{2}{\epsilon(2\alpha+2\beta+1)}\exp\left\{\frac{\epsilon}{2}\left[\ln4\pi-\gamma-\frac{\epsilon}{4}\zeta_2\right]\right\}\nn
&\times\exp\left\{\sum_{j=1}^{\infty}(\alpha+\beta+1)^j\frac{\epsilon^j}{j}+\sum_{j=3}^{\infty}h_j(\alpha,\beta)\zeta_j\frac{\epsilon^j}{j}\right\}
\end{align}
where
\be
h_j(\alpha,\beta)=(\alpha+\tfrac12)^j+(\beta+\tfrac12)^j+(-\alpha-\beta-\tfrac12)^j-(-\alpha)^j-(-\beta)^j-(\alpha+\beta+1)^j.
\ee
It is easy to check that
\be
h_4=\frac34h_3+\frac{1}{64},
\ee
irrespective of $\alpha$, $\beta$; and there are similar expressions for $h_j$ for higher even values of $ j$. 
 We would like to use this to derive a relation between $ F_{L,m}^{\zeta_4}$ and $ F_{L,m+1}^{\zeta_3}$ which we will be able to exploit in Eqs.~\eqref{betas}, \eqref{double}; but these quantities may include $ \zeta$-dependent counterterms which do not obey this  relation between $ h_4$ and $ h_3$.
Accordingly, we define $ G^{\zeta_n}_{L,m}$ to be the value of $ F^{\zeta_n}_{L,m}$ after omitting $ \zeta_n$-dependent counterterms.
Then we do have 
\be
 G^{\zeta_4}_{L,m}=\frac34G^{\zeta_3}_{L,m+1} \quad (m\ge0).
\label{Grela}
\ee 
We further obtain
\begin{align}
G_{L,m}^{\zeta_6}=&\frac54\left(G_{L,m+1}^{\zeta_5}-\frac13G_{L,m+2}^{\zeta_4}\right),\nn
G_{L,m}^{\zeta_8}=&\frac74\left(G_{L,m+1}^{\zeta_7}-\frac12G_{L,m+2}^{\zeta_6}+\frac{1}{24}G_{L,m+4}^{\zeta_4}\right).
\label{Grelb}
\end{align}
Relations like this can be extended to higher even $ \zeta$ and higher loops (in fact for almost\footnote{See the Conclusions for a clarification  of this {\it caveat}.} all known ``$ p$-integrals'' - two-point integrals with a single momentum dependence). Consequently, the $G_{L,m}^{\zeta_k}$ for even $k$ may be recursively written in terms of the $G_{L,m}^{\zeta_k}$ for lower odd values of $k$.
\section{Scheme redefinitions}
In this Section we shall show how relations such as those in Eqs.~\eqref{Grela} and \eqref{Grelb} may be exploited to define scheme redefinitions making a transformation from $\msbar$ to a scheme in which even $\zeta$s are absent.

We have from Eq.~\eqref{Grela} combined with the definition of the $G^{\zeta_n}_{l,m}$, 
\begin{align}
G^{\zeta_3}_{3,1}=F^{\zeta_3}_{3,1},\quad G^{\zeta_4}_{3,0}=&F^{\zeta_4}_{3,0},\quad  G^{\zeta_4}_{4,1}=F^{\zeta_4}_{4,1},\nn
G^{\zeta_3}_{4,2}=&F^{\zeta_3}_{4,2}-F^{\zeta_3}_{3,1}\cdot F_{1,1}.
\end{align}
Combining this with Eq.~\eqref{double}, we obtain
\be
G^{\zeta_3}_{4,2}=\frac14[F_{1,1},F^{\zeta_3}_{3,1}]=\frac14[\beta_1,G^{\zeta_3}_{3,1}].
\ee
Using 
\begin{align}
G^{\zeta_4}_{4,1}=&\frac34G^{\zeta_3}_{4,2},\nn
G^{\zeta_4}_{3,0}=&\frac34G^{\zeta_3}_{3,1}.
\end{align}
we then find that 
\be
\beta^{\zeta_4}_4=4F^{\zeta_4}_{4,1}=4G^{\zeta_4}_{4,1}=[\beta_1,G^{\zeta_4}_{3,0}]=[\beta_1,F^{\zeta_4}_{3,0}].
\ee
In the light of Eq.~\eqref{delb}, this may be removed by a coupling redefinition with
\be
\delta g=-F^{\zeta_4}_{3,0}\zeta_4
\label{Fsub}
\ee
i.e. removing the $ \zeta_4$-dependent finite part of $F$. 
 This idea may clearly be extended to higher orders; using the further relations in Eq.~\eqref{Grelb}, our proposal is that  all even $ \zeta$ may be removed by 
\be
\delta g= -\Bigl(\left[F_{3,0}^{\zeta_4}+F_{4,0}^{\zeta_4}\right]\zeta_4
+\left[F_{4,0}^{\zeta_6}+F_{5,0}^{\zeta_6}\right]\zeta_6
+\left[F_{5,0}^{\zeta_8}+F_{6,0}^{\zeta_8}\right]\zeta_8\Bigr)+\ldots
\ee
This is a scheme where we subtract all even-$ \zeta$-dependent finite parts in $F$- we call it $\hbox{MOM}^{\prime}$. We have shown cancellation of even $\zeta$s in the \momp scheme up to a loop order beyond their first appearance for $ \zeta_4$ (i.e. 5 loops) and $ \zeta_6$ (i.e. 6 loops). 
We have also examined the MOM scheme in which we subtract {\it  all} finite parts. As we have mentioned, there is a considerable literature devoted to this scheme, though largely in the single-coupling case. We have shown (at least up to five loops for $ \zeta_4$) that even $ \zeta$s also cancel in MOM for a general multi-coupling theory; it seems likely that MOM shares the no-$\pi$ property with \momp at higher orders as well.

\section{ Example: the Wess-Zumino model}
The supersymmetric Wess-Zumino model is naturally written in terms of superfields $\Phi$ with a cubic superpotential given by Eq.~\eqref{Wdef}. As we have suggested already, the Wess-Zumino model is a useful test-bed since the $\beta$-function is determined by the anomalous dimensions. Schematically,
\be
 \beta=\Scal_3 \tikz[baseline=(vert_cent.base)]{
  \node (vert_cent) {\hspace{-13pt}$\phantom{-}$};
 \draw (-1.4,0.3)--(-1,0);
\draw (-1.4,-0.3)--(-1,0);
\draw (-1,0)--(-0.6,0);
 \draw (-0.3,0) circle [radius=0.3cm];
\draw (0,0)--(0.3,0);
\draw (-0.3,0) node {$\gamma$}
        }
\label{bdef}
\ee
where $ \Scal_3$ denotes the sum over the three terms where $ \gamma$ is attached to each external line.
Likewise we consider variations $\delta g$ given by
\be
 \delta g=\Scal_3 \tikz[baseline=(vert_cent.base)]{
  \node (vert_cent) {\hspace{-13pt}$\phantom{-}$};
 \draw (-1.4,0.3)--(-1,0);
\draw (-1.4,-0.3)--(-1,0);
\draw (-1,0)--(-0.6,0);
 \draw (-0.3,0) circle [radius=0.3cm];
\draw (0,0)--(0.3,0);
\draw (-0.3,0) node {$h$}
        }.
\label{dgdef}
\ee
The anomalous dimension is given up to four loops by
\begin{align}
\gamma= &\frac12 \, 
\tikz[baseline=(vert_cent.base)]{
  \node (vert_cent) {\hspace{-13pt}$\phantom{-}$};
  \draw (-0.3,0)--(0.1,0);
     \draw   (0.7,0) ++(0:0.6cm) arc (0:360:0.6cm and 0.4cm)
        (1.3,0)--(1.7,0);
        \filldraw [gray] (0.1,0) circle [radius=1.5pt];
}+\ldots-\frac34(2\zeta_3+\zeta_4)\, 
\tikz[baseline=(vert_cent.base)]{
  \node (vert_cent) {\hspace{-13pt}$\phantom{-}$};
  \draw (-0.3,0)--(-0.1,0)
     (0.7,0) ++(0:0.8cm and 0.5cm) arc (0:45:0.8cm and 0.5cm) node (n1)  {}
            (0.7,0) ++(135:0.8cm and 0.5cm) arc (135:360:0.8cm and 0.5cm) 
                (0.7,0.36) ++(0:0.55cm and 0.35cm) arc (0:50:0.55cm and 0.35cm) node (n2) {}
          (0.7,0.36) ++(50:0.55cm and 0.35cm) arc (50:130:0.55cm and 0.35cm)  node (n3)  {}
           (0.7,0.36) ++(130:0.55cm and 0.35cm) arc (130:230:0.55cm and 0.35cm) node (n4)  {}
            (0.7,0.36) ++(230:0.55cm and 0.35cm) arc (230:310:0.55cm and 0.35cm) node (n5)  {}
             (0.7,0.36) ++(310:0.55cm and 0.35cm) arc (310:360:0.55cm and 0.35cm) node (n6)  {}
             (n2.base) to (0.74,0.4)
             (0.62,0.3) to (n4.base)
             (n3.base) to (n5.base)
         (1.5,0)--(1.7,0);
                   \filldraw [gray] (n1) circle [radius=1.5pt]; 
         \filldraw [gray]  (-0.1,0) circle [radius=1.5pt];
         \filldraw [gray] (n3) circle [radius=1.5pt]; 
          \filldraw [gray] (n4) circle [radius=1.5pt];
          }\nn
-\frac34&(2\zeta_3-\zeta_4)\left(
 \tikz[baseline=(vert_cent.base)]{
  \node (vert_cent) {\hspace{-13pt}$\phantom{-}$};
  \draw[red] (-0.3,0)--(-0.1,0);\draw
     (0.7,0) ++(0:0.8cm and 0.5cm) arc (0:45:0.8cm and 0.5cm) node (n1)  {};\draw
     (0.7,0) ++(45:0.8cm and 0.5cm) arc (45:95:0.8cm and 0.5cm) node(n2) {};\draw[red]
              (0.7,0) ++(95:0.8cm and 0.5cm) arc (95:115:0.8cm and 0.5cm)  node(n3) {} ;\draw[red]
          (0.7,0) ++(115:0.8cm and 0.5cm) arc (115:155:0.8cm and 0.5cm) node(n4) {}
            (0.7,0) ++(155:0.8cm and 0.5cm) arc (155:265:0.8cm and 0.5cm)  node(n5) {};\draw
             (0.7,0) ++(265:0.8cm and 0.5cm) arc (265:315:0.8cm and 0.5cm) node(n6) {}
              (0.7,0) ++(315:0.8cm and 0.5cm) arc (315:360:0.8cm and 0.5cm)
            (n1.base) to (1.03,0.05)
            (0.95,-0.05) to (n5.base)
           (n2.base) to (n6.base); \draw[red]
     (n3.base) to  [out= 320,in= 300] (n4.base) ;\draw
         (1.5,0)--(1.7,0);
                   \filldraw [gray] (n1) circle [radius=1.5pt]; 
          \filldraw [gray]  (-0.1,0) circle [radius=1.5pt];
         \filldraw [gray] (n3) circle [radius=1.5pt]; 
         \filldraw [gray] (n6) circle [radius=1.5pt]; 
          }
          + 
 \tikz[baseline=(vert_cent.base)]{
  \node (vert_cent) {\hspace{-13pt}$\phantom{-}$};
  \draw (-0.3,0)--(-0.1,0);
   \draw[red]  (0.7,0) ++(0:0.8cm and 0.5cm) arc (0:25:0.8cm and 0.5cm) node (n1)
         {}
     (0.7,0) ++(25:0.8cm and 0.5cm) arc (25:65:0.8cm and 0.5cm) node(n2)
       {};\draw[red]
        (0.7,0) ++(65:0.8cm and 0.5cm) arc (65:85:0.8cm and 0.5cm) node(n3){};\draw[red]
        (0.7,0) ++(85:0.8cm and 0.5cm) arc (85:135:0.8cm and 0.5cm) node(n4){};\draw
        (0.7,0) ++(135:0.8cm and 0.5cm) arc (135:225:0.8cm and 0.5cm)  node(n5) {} 
        (0.7,0) ++(225:0.8cm and 0.5cm) arc (225:275:0.8cm and 0.5cm) node(n6) {}
            (0.7,0) ++(275:0.8cm and 0.5cm) arc (275:360:0.8cm and 0.5cm);
            \draw[red] (n1.base) to  [out=240,in=240] (n2.base) ;\draw[red]
             (n3.base) to  (0.43,0.05)
             (0.37,-0.05) to (n5.base) ;\draw
     (n4.base) to  (n6.base) 
         (1.5,0)--(1.7,0);
                   \filldraw [gray] (n1) circle [radius=1.5pt]; 
          \filldraw [gray]  (-0.1,0) circle [radius=1.5pt];
          \filldraw [gray] (n3) circle [radius=1.5pt]; 
           \filldraw [gray] (n6) circle [radius=1.5pt]; 
  }       
+  2\,
  \tikz[baseline=(vert_cent.base)]{
  \node (vert_cent) {\hspace{-13pt}$\phantom{-}$};
  \draw (-0.3,0)--(-0.1,0);\draw[red]
     (0.7,0) ++(0:0.8cm and 0.5cm) arc (0:45:0.8cm and 0.5cm) node (n1)  {}
     (0.7,0) ++(45:0.8cm and 0.5cm) arc (45:73:0.8cm and 0.5cm) node(n5)  {}
     (0.7,0) ++(73:0.8cm and 0.5cm) arc (73:107:0.8cm and 0.5cm) node(n6)  {}
     (0.7,0) ++(107:0.8cm and 0.5cm) arc (107:135:0.8cm and 0.5cm) node(n2)  {};\draw
        (0.7,0) ++(135:0.8cm and 0.5cm) arc (135:225:0.8cm and 0.5cm) node(n3){}
          (0.7,0) ++(225:0.8cm and 0.5cm) arc (225:315:0.8cm and 0.5cm)  node(n4) {} 
          (0.7,0) ++(315:0.8cm and 0.5cm) arc (315:360:0.8cm and 0.5cm);\draw[red]
           (n1.base) to (0.76,0.05);\draw
         (n2.base) to  (n4.base) ;\draw[red]
         (0.65,-0.05) to (n3.base);\draw[red]
         (n5.base) to [out = 280,in = 260] (n6.base);\draw
         (1.5,0)--(1.7,0);
                   \filldraw [gray] (n1) circle [radius=1.5pt]; 
          \filldraw [gray]  (-0.1,0) circle [radius=1.5pt];
          \filldraw [gray] (n4) circle [radius=1.5pt]; 
           \filldraw [gray] (n6) circle [radius=1.5pt];
          }\right)+\ldots.
\label{gam}
\end{align}
Here, in contrast to Eq.~\eqref{diag}, each diagram denotes purely a combination of contracted tensor couplings and the results of Feynman integrals are subsumed into the coefficients multiplying the diagrams. The dots at vertices denote the complex conjugated terms in Eq.~\eqref{Wdef}; since in this theory the propagators link a $\Phi$ to a $\bar\Phi$, the contractions are always between conjugated and unconjugated tensors. We only show the terms relevant for our discussion, omitting all the two-and three-loop terms and the non-$\zeta_4$-dependent four-loop terms. In line with Eq.~\eqref{Fsub} and in the light of Eq.~\eqref{dgdef}, we have 
\be
h^{(3)}=-\frac14\gamma^{\zeta_3}_3\zeta_4
=-\frac38\zeta_4\tikz[baseline=(vert_cent.base)]{
  \node (vert_cent) {\hspace{-13pt}$\phantom{-}$};
  \draw (-0.4,0)--(0.1,0)
     (0.7,0) ++(0:0.6cm and 0.4cm) arc (0:50:0.6cm and 0.4cm) node (n1)
            {}
     (0.7,0) ++(50:0.6cm and 0.4cm) arc (50:130:0.6cm and 0.4cm) node(n2)
       {}
        (0.7,0) ++(130:0.6cm and 0.4cm) arc (130:230:0.6cm and 0.4cm) node(n3){}
          (0.7,0) ++(230:0.6cm and 0.4cm) arc (230:310:0.6cm and 0.4cm)  node(n4) {} 
          (0.7,0) ++(310:0.6cm and 0.4cm) arc (310:360:0.6cm and 0.4cm) 
           (n1.base) to (0.75,0.05)
         (n2.base) to  (n4.base) 
         (0.65,-0.05) to (n3.base)
         (1.3,0)--(1.8,0);
                   \filldraw [gray] (n1) circle [radius=1.5pt]; 
          \filldraw [gray]  (0.1,0) circle [radius=1.5pt];
          \filldraw [gray] (n4) circle [radius=1.5pt]; 
          }.
\label{hdef}
\ee
Eq.~\eqref{delb} then leads to (once again using Eq.~\eqref{dgdef})
\begin{align}
\delta\beta=&[\delta g+\delta g^*,\beta]\imp \delta \gamma=(\delta g+\delta g^*)\cdot\gamma-(\beta+\beta^*)\cdot h\nn
=&(\delta g)^{(3)klm}\frac{\pa}{\pa g^{klm}}\gamma^{(1)}
-\beta^{(1)klm}\frac{\pa}{\pa g^{klm}}h^{(3)}+*\hbox{ terms}.
\label{dbWZ}
\end{align}
It is then easy to check using Eqs.~\eqref{bdef}, \eqref{gam} and \eqref{hdef} that the $ \zeta_4$ terms cancel. We have indicated in red the terms in Eq.~\eqref{gam} which form up into an insertion of $\beta^{(1)}$ on $h^{(3)}$ and are therefore cancelled by the second term in Eq.~\eqref{dbWZ}.
\section{Conclusions}
Our goal in this talk has been to seek renormalisation schemes in which even $\zeta$-functions are absent from renormalisation group functions for a general theory. To this end we have considered two renormalisation schemes, MOM and $\hbox{MOM}^{\prime}$. The MOM scheme is defined by subtracting $O(\epsilon^0)$ parts of $n$-point functions in addition to poles in $\epsilon$, while in the \momp scheme we only subtract even-$\zeta$ finite parts. We have considered a general multi-coupling theory, focussing attention on one where the $\beta$-function is defined by the anomalous dimensions; the supersymmetric Wess Zumino model is a concrete example. In this context we have shown (for more details see Ref.~\cite{jack}) that the no-$\pi$ theorem holds for $ \zeta_4$   up 5 loops  in both the \momp and MOM schemes and for $ \zeta_6$ up to 6 loops in the \momp scheme. We believe that these results may be extended to higher loops and higher even-$\zeta$s for both MOM and $\hbox{MOM}^{\prime}$, but evidently this will require more work. The theoretical underpinning of our work is based on $ p$-integrals, i.e. integrals with a single momentum dependence; these naturally arise in two-point functions, hence our primary interest in theories whose renormalisation is determined by anomalous dimensions. However we believe that an extension to 3-point and 4-point graphs is feasible, based on nullification of one or two (respectively) of the external momenta. There are certainly well-motivated nullification procedures for 3-point vertices  in QCD, as described in Ref.~\cite{gracey3}. Furthermore this nullification procedure has also been carried out for $\phi^3$ theory in six dimensions\cite{gracey5}. The implementation of a MOM-type scheme in both these cases leads to the expected absence of even $\zeta$s up to the loop order considered. It seems likely that we shall be able to apply the same procedure to $\phi^4$ theory in four dimensions, despite additional complications arising in this case from potential infra-red issues when nullifying momenta in 4-point graphs. Finally there is evidence that the expected behaviour of $p$-integrals on which we have been relying breaks down at high loop orders\cite{loopseven}\footnote{We are grateful to David Broadhurst and Oliver Schnetz for discussions on this point.}-in fact at eight loops, for $\zeta_{12}$.

\acknowledgments

Financial support to attend this conference was provided by the STFC Consolidated Grant ST/T000988/1.

\end{document}